\documentclass[manuscript]{aastex}

\def\erg{\hbox{\rm erg}}
\def\ergs{\hbox{\rm erg$\,$s$^{-1}$}}
\def\kms{\hbox{km$\,$s$^{-1}$}}
\def\VLSR{\hbox{$V_{\rm LSR}$}}

\def\sun{\hbox{$\odot$}}

\def\lesssim{\mathrel{\hbox{\rlap{\hbox{\lower4pt\hbox{$\sim$}}}\hbox{$<$}}}}
\def\gtrsim{\mathrel{\hbox{\rlap{\hbox{\lower4pt\hbox{$\sim$}}}\hbox{$>$}}}}

\def\arcdeg{\hbox{$^\circ$}}
\def\arcmin{\hbox{$^\prime$}}
\def\arcsec{\hbox{$^{\prime\prime}$}}

\slugcomment{To appear in the Astrophysical Journal Letters}
\shorttitle{Kinematics of the Ultra-High-Velocity Gas}
\shortauthors{Yamada et al.}

\begin{document}
\title{Kinematics of Ultra-High-Velocity Gas in the Expanding Molecular Shell \\ adjacent to the \object{W44} Supernova Remnant}

\author{Masaya Yamada\altaffilmark{1}, Tomoharu Oka\altaffilmark{1,2}, Shunya Takekawa\altaffilmark{1}, Yuhei Iwata\altaffilmark{1}, Shiho Tsujimoto\altaffilmark{1}, Sekito Tokuyama\altaffilmark{1}, Maiko Furusawa\altaffilmark{1}, Keisuke Tanabe\altaffilmark{1}, \&\ Mariko Nomura\altaffilmark{2} }
\affil{\altaffilmark{1}School of Fundamental Science and Technology, Graduate School of Science and Technology, Keio University, 3-14-1 Hiyoshi, Kohoku-ku, Yokohama, Kanagawa 223-8522, Japan}
\affil{\altaffilmark{2}Department of Physics, Faculty of Science and Technology, Keio University, 3-14-1 Hiyoshi, Kohoku-ku, Yokohama, Kanagawa 223-8522, Japan}
\email{yamada@aysheaia.phys.keio.ac.jp}

\begin{abstract}
We mapped the ultra-high-velocity feature (the ``Bullet'') detected in the expanding molecular shell associated with the \object{W44} supernova remnant using the Nobeyama Radio Observatory 45-m telescope and the ASTE 10-m telescope.  The Bullet clearly appears in the CO {\it J}=1--0, CO {\it J}=3--2, CO {\it J}=4--3, and HCO$^+$ {\it J}=1--0 maps with a compact appearance ($0.5\times 0.8$ pc$^2$) and an extremely broad velocity width ($\Delta V\!\simeq\!100 \ \mbox{\kms}$).  The line intensities indicate that the Bullet has a higher density and temperature than those in the expanding molecular shell.  The kinetic energy of the Bullet amounts to $10^{48.0}  \ \mbox{\erg}$ which is approximately 1.5 orders of magnitude greater than the kinetic energy shared to the small solid angle of it.  Two possible formation scenarios with an inactive isolated black hole (BH) are presented.
\end{abstract}

\section{INTRODUCTION}
W44 is a ``mixed-morphology'' supernova remnant (SNR) with a radio continuum shell filled with a center-peaked X-ray emission (e.g., Gronenschild et al. 1978; Smith et al. 1985; Rho et al. 1994).  It has a radio pulsar within the boundary of the shell (Wolszczan et al. 1991). The distance to the SNR is considered to be 3 kpc based on HI absorption studies (Caswell et al. 1975).  The W44 SNR interacts with the adjacent giant molecular cloud (Seta et al. 1998, 2004).  Shocked molecular gas is well traced by 25 OH 1720 MHz maser spots (Claussen et al. 1997), several high-velocity wings of CO emission (Seta et al. 2004), and by a faint spatially-extended broad CO and HCO$^+$ emissions (Sashida et al. 2013). Such OH maser and broad emissions arise from a thin expanding shell of shocked molecular gas with the expansion velocity of $13.2 \ \mbox{\kms}$ (Sashida et al. 2013).  The total kinetic energy, $E_{\rm kin}=(1\mbox{--}3)\!\times\!10^{50}  \ \mbox{\erg}$, corresponds to the yield from a supernova explosion of $0.1\mbox{--}0.3$.  

In the process of investigating the kinematics of the W44 expanding shell, we noticed an extraordinarily broad velocity feature at $(l, b)\!=\!(34.725\arcdeg, -0.472\arcdeg )$ in CO {\it J}=3--2 data (Fig.1b).  The full-width-zero-intensity (FWZI) velocity width of this feature exceeds $100 \ \mbox{\kms}$, while those of the previously detected broad emissions are less than $30 \ \mbox{\kms}$.  The position of this ultra-high-velocity feature is $\sim\!0.15\arcdeg$ ($\sim\!8$ pc in projected distance) apart from the W44 radio pulsar (Wolszczan et al. 1991).  
By its appearance in position-velocity maps, we named this ultra-high-velocity feature the ``Bullet''.  
The kinematics of the Bullet differ distinctly from those of the expanding molecular shell.  Interestingly, a radio continuum blob (Jones et al. 1993) and an H$_2$ ro-vibrational line nebulosity (Reach et al. 2005) overlap with the Bullet in the plane of the sky (Sashida et al. 2013).  Despite the high significance of the Bullet in the CO {\it J}=3--2 data, neither its nature nor origin have been totally explained.  To determine the origin of the Bullet, we performed detailed mappings in several millimeter and submillimeter lines.  Throughout this study, the distance to the Bullet (W44 SNR) is assumed to be 3 kpc.  

%%%%%%%    Place Fig.1 here, please.      %%%%%%% 
\begin{figure*}[htbp]
\epsscale{0.87}
\plotone{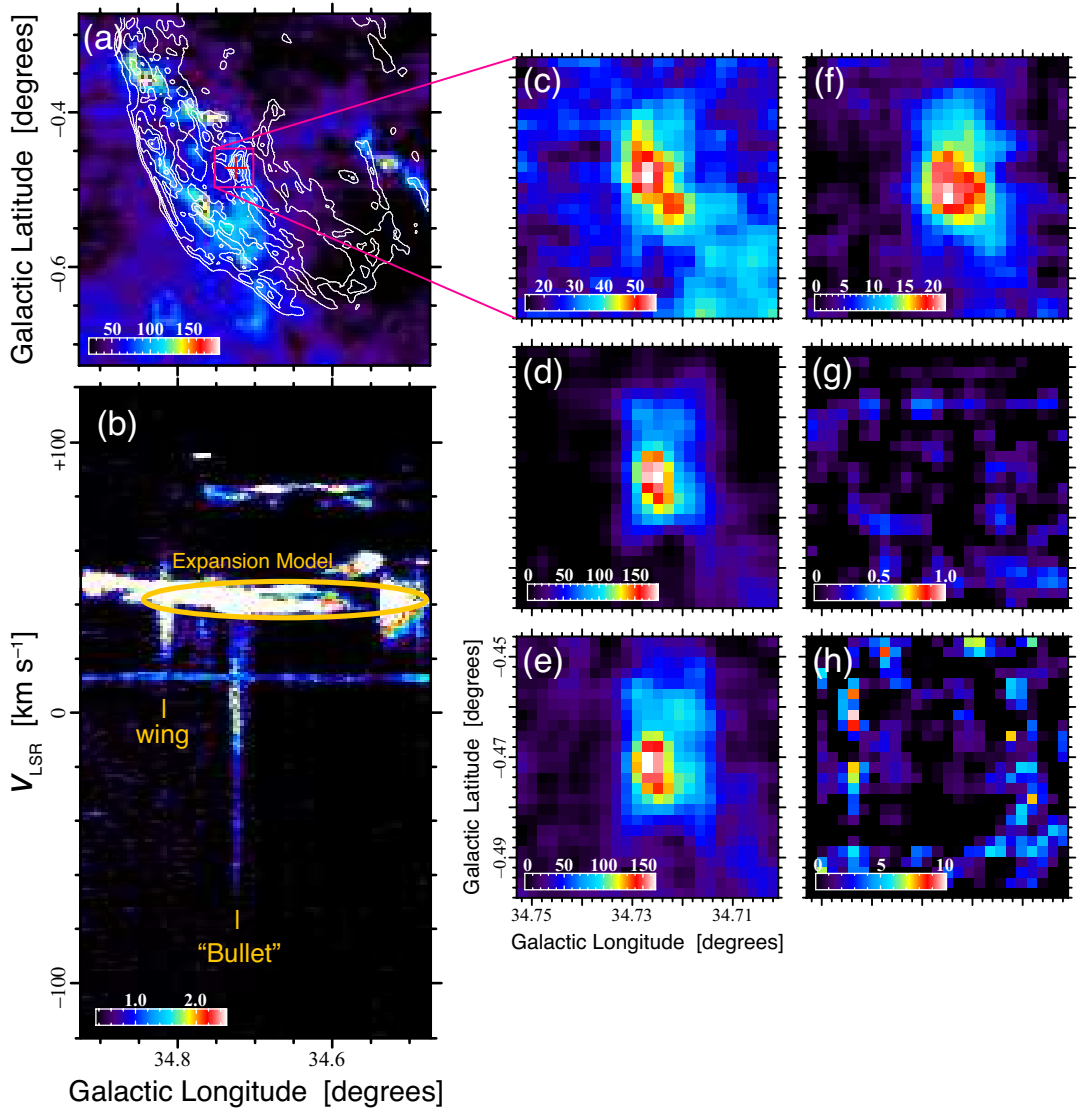}
\caption{(a) Map of the velocity-integrated CO {\it J}=3--2 emission (Sashida et al. 2013). The velocity range for the integration is from $\VLSR=+20 \ \mbox{\kms}$ to $+60 \ \mbox{\kms}$. White contours outline VLA radio continuum emission at 20 cm (Jones et al. 1993). The orange ellipse shows the best-fit model of uniformly expanding shell calculated from the spatially-extended wings(Sashida et al. $2013$). The magenta rectangle indicates the mapping area of the new observations. The red cross marks the position of the Bullet. (b) Longitude--velocity map of CO {\it J}=3--2 emission at $b=-0.472\arcdeg$ (Sashida et al. 2013).
Velocity-integrated maps of (c) CO {\it J}=1--0, (d) CO {\it J}=3--2, (e) CO {\it J}=4--3, (f) HCO$^+$ {\it J}=1--0, (g) HCO$^+$ {\it J}=4--3, and (h) C$^0$ $^3P_1$--$^3P_0$ emissions. The velocity range for the integration is $\VLSR\!=\!-79$--$+34 \ \mbox{\kms}$. All intensities are presented in the $T_{\rm MB}$ scale.
\label{fig1}}
\end{figure*}  
%%%%%%%%%%%%%%%%%%%%%%%%%%%%%%

\section{OBSERVATIONS}
\subsection{ASTE}
We mapped the Bullet in the CO {\it J}=3--2 ($345.796$ GHz), HCO$^+$ {\it J}=4--3 ($356.734 \,\mbox{GHz}$), CO {\it J}=4--3 ($461.041 \,\mbox{GHz}$), and C$^0$ $^3P_1$--$^3P_0$ ($492.161 \,\mbox{GHz}$) lines using the Atacama Submillimeter Telescope Experiment (ASTE) 10-m telescope.  For comparison, we also mapped Wing 1 (Seta et al. 2004) as a representative of shocked molecular gas adjacent to the W44 SNR.  The observations were conducted on July 30, August 5--8, and 25--26 in 2014, as well as July 21--24, 28, September 12--13, and 21--24 in 2015. The main beam efficiency ($\eta_{\rm MB}$) and FWHM beamsize of the telescope are $0.6$ and $22\arcsec$ at $350 \,\mbox{GHz}$, $0.45$ and $17{\arcsec}$ at $492 \,\mbox{GHz}$, respectively. The pointing of the telescope was checked and corrected every 2 h by observing CO emission from a late-type star R Aql, and its accuracy was maintained within 2${\arcsec}$ (rms). The observations were made with a cartridge-type single-polarization side-band separating (2SB) mixer receiver CATS345 in 2014, and a 2SB mixer receiver DASH345 and dual-polarization 2SB mixer receiver ASTE BAND8 in 2015. All observations were made in OTF mapping mode. The reference position was taken at $(l, b)\!=\!(+33.750\arcdeg, -1.509\arcdeg)$. During the observations, the system noise temperatures ranged between 200 and 600 K (SSB) with CATS345 and DASH345, 600 and 1500 K (SSB) with ASTE BAND8. The antenna temperature (${T_{\rm A}}^{\ast}$) was obtained by the standard chopper-wheel technique.

We used an XF-type digital spectro-correlator MAC in the 512 MHz bandwidth (0.5 MHz resolution) mode.
This combination of the frequency bandwidth and resolution corresponds to $444 \ \mbox{\kms}$\ velocity coverage with a $0.43 \ \mbox{\kms}$\ velocity resolution at $346 \,\mbox{GHz}$ and $333  \ \mbox{\kms}$\ coverage with a $0.33 \ \mbox{\kms}$\ resolution at $461 \,\mbox{GHz}$. The obtained data were reduced using the NOSTAR reduction package. We subtracted the baselines of all spectra by fitting first- or third-order polynomial lines. We scaled the antenna temperature by multiplying it by $1 / \eta_{\rm MB}$ to obtain the main-beam temperature, $T_{\rm MB}$. The data were smoothed with a Bessel--Gaussian function and resampled on an $7.5\arcsec\!\times\!7.5\arcsec\!\times\!1.0 \ \mbox{\kms}$ grid to obtain the final maps.

%%%%%%%    Place Fig.2 here, please.      %%%%%%% 
\begin{figure*}[htbp]
\centering
\epsscale{0.85}
\plotone{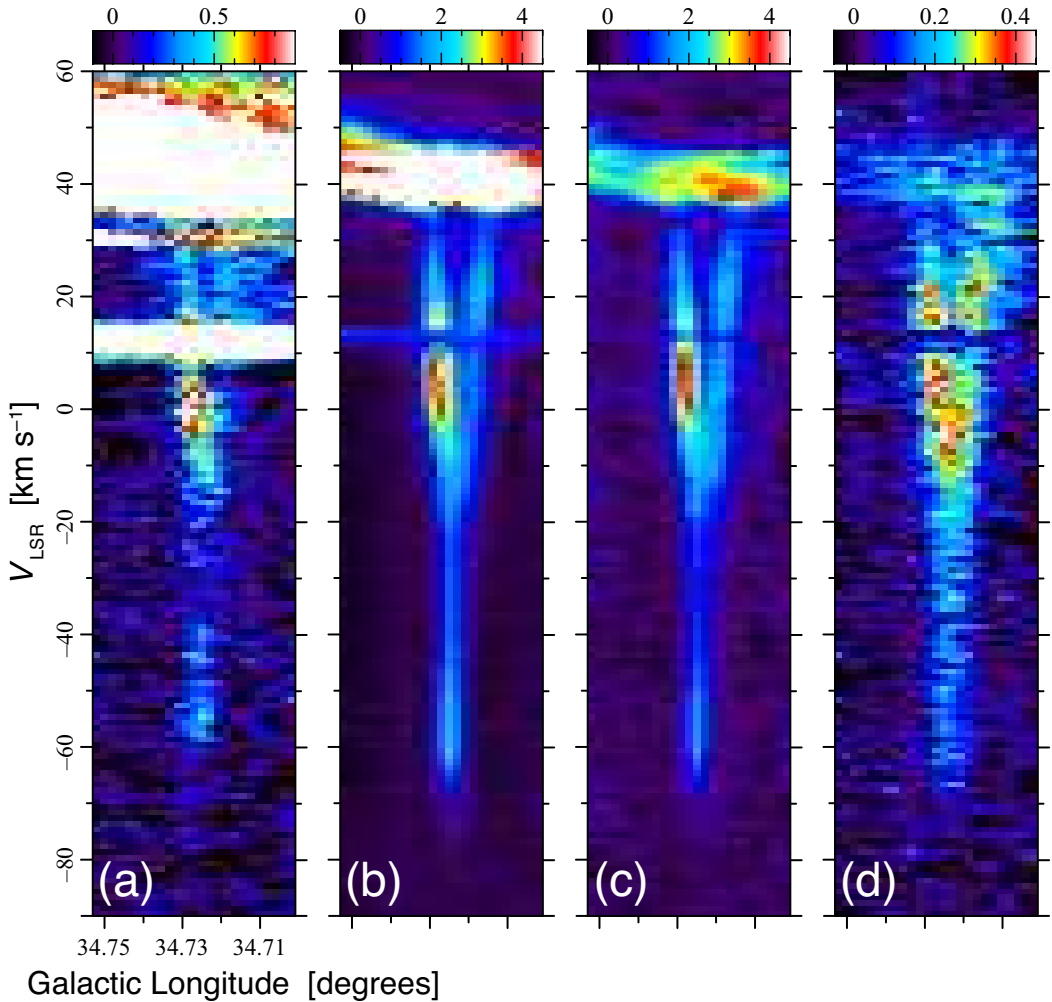}
\caption{Longitude--velocity maps of (a) CO {\it J}=1--0, (b) CO {\it J}=3--2, (c) CO {\it J}=4--3, and (d) HCO$^+$ {\it J}=1--0 emissions at $b$ = $-$0.472\arcdeg.
All intensities are presented in the $T_{\rm MB}$ scale. 
\label{fig2}}
\end{figure*}
%%%%%%%%%%%%%%%%%%%%%%%%%%%%%%

\subsection{NRO 45-m}
The Bullet and Wing 1 were also mapped in the CO {\it J}=1--0 ($115.271 \,\mbox{GHz}$) and HCO$^+$ {\it J}=1--0 ($89.1885 \,\mbox{GHz}$) lines with the Nobeyama Radio Observatory (NRO) 45-m radio telescope. The NRO 45-m observations were made in 2015 January 8--15 and March 5--7. The main beam efficiencies and FWHM beam sizes of the telescope are $0.5$ and $20\arcsec$ at $86 \,\mbox{GHz}$, $0.4$ and $15\arcsec$ at $115 \,\mbox{GHz}$, respectively. The telescope pointing was checked and corrected every 2 h by observing the SiO maser source V1418 Aql at $43 \,\mbox{GHz}$, and its accuracy was maintained within 2$\arcsec$ (rms). All data were obtained with 2SB receivers TZ1V/H in OTF mapping mode. The reference position was taken at $(l, b)\!=\!(+33.750\arcdeg, -1.509\arcdeg)$. During the observations, the system noise temperatures ranged between 100 and 400 K (SSB) with TZ1V/H. The antenna temperature was calibrated by the standard chopper-wheel technique.

We used the highly flexible FX-type correlator SAM45 in the 500 MHz bandwidth (122 kHz resolution) mode.
This combination of the frequency bandwidth and resolution corresponds to $1300 \ \mbox{\kms}$ velocity coverage with a $0.32 \ \mbox{\kms}$ velocity resolution at $115 \,\mbox{GHz}$, and $1700 \ \mbox{\kms}$ coverage with a $0.41 \ \mbox{\kms}$ resolution at $89\,\mbox{GHz}$, respectively.  
The NRO 45-m data were reduced using the NOSTAR reduction package, and the final maps were generated with the same parameters as those of the ASTE maps.

\section{RESULTS and DISCUSSION}
\subsection{Spatial-Velocity Structure}
Figures 1c to 1h show the velocity-integrated maps of the $3\arcmin\times 3\arcmin$ ($2.6\times 2.6$ pc$^2$) area around the Bullet in the six observed lines. The Bullet appears as a compact clump with a size of $0.5\times 0.8$ pc$^2$ in the CO {\it J}=1--0, CO {\it J}=3--2, CO {\it J}=4--3, and HCO$^+$ {\it J}=1--0 maps. The detected line emissions peak at $(l, b)\!\simeq\!(34.73\arcdeg, -0.47\arcdeg)$. The entity of the Bullet is well-defined by CO {\it J}=3--2 and \mbox{CO {\it J}=4--3} emissions, while C$^0$ $^3P_1$--$^3P_0$ emission is absent. A blob of 20 cm continuum emission traces the Galactic eastern rim of the Bullet (Fig. 1a). A nebulosity of H$_2$ ro-vibrational transition line emission overlaps the Galactic eastern half of the Bullet (Reach et al. 2005). An inspection of the infrared (Yamauchi et al. 2011) and X-ray (Muno et al. 2006) point source catalogs reveals that no counterpart of the Bullet is detected in those wavelengths.  

Longitude--velocity maps illustrate the broad-velocity nature of the Bullet (Fig. 2). Its velocity width in FWZI well exceeds $100 \ \mbox{\kms}$. The narrow velocity width features at $\VLSR\!\simeq\!12 \ \mbox{\kms}$\ and $30\ \mbox{\kms}$\ are Galactic spiral arms irrelevant to the W44 molecular cloud. The Bullet is very compact ($\sim\!0.8$ pc) in its negative high-velocity half ($\VLSR\!\leq\!-10 \ \mbox{\kms}$ ). It bifurcates in the lower velocity range ($\VLSR\!\geq\!-10 \ \mbox{\kms}$ ), exhibiting a ``Y'' shape, and merges into the W44 molecular cloud at $\VLSR\!\simeq\!40 \ \mbox{\kms}$. CO {\it J}=3--2, CO {\it J}=4--3, and HCO$^+$ {\it J}=1--0 emissions decline at the footpoint of the Bullet ($\VLSR\!\sim\!+30 \ \mbox{\kms}$ ).

\subsection{Physical Parameters}
We estimated the physical conditions in the Bullet and Wing 1 by employing a large velocity gradient (LVG) model calculations (Goldreich \&\ Kwan 1974).  Table 1 summarizes the line intensities and the best-fit value of physical conditions. Beam-filling factors were assumed to be unity.  We used the main-beam temperatures at $(l, b, \VLSR)\!=\!(34.725\arcdeg, -0.472\arcdeg, -57.5\,\mbox{\kms})$ and $(l, b, \VLSR)\!=\!(34.754\arcdeg, -0.406\arcdeg, +36.8\,\mbox{\kms})$ as representative positions of the Bullet and Wing 1, respectively. We assumed the fractional abundance [CO]/[H$_2$]$\!=10^{-4.1}$ (Frerking, Langer, \&\ Wilson 1982).  
The 90\%\ confidence intervals are also shown in Table 1.  We see that the Bullet clearly has a higher temperature and lower density than Wing 1, which represent shocked gas in the W44 molecular cloud.  

%%%%%%%    Place Table 1 here, please.      %%%%%%%%%%%%%%%%%%%%%%%%%%%%%%%%%%%%%%%%%%%%%%%%%%%%%%%%%%%%%%%%%%%%%
%\renewcommand{\arraystretch}{2}
\begin{table*}[]
%\begin{center}
\centering
\begin{minipage}[]{43.5em}%{12.5 cm}
\caption{LVG model calculations of Wing 1 and Bullet}
\begin{tabular}{ccccccc}
\hline \hline
Source   &   &$T_{\rm MB}$ [K]   &   &$T_{\rm k}$   &${\rm log}(n(\rm H_2))$   &${\rm log}( N({\rm H_2})/ \Delta V )$
\rule[-1.5mm]{0mm}{4.5mm} \\
   &CO {\it J}=1--0   &CO {\it J}=3--2   &CO {\it J}=4--3   &[K]   &$\rm [cm^{-3}]$   
&$[\rm cm^{-2} (\kms)^{-1}]$ 
\rule[-1.5mm]{0mm}{4.5mm} \\ \hline 
Bullet   &$0.40\pm 0.07$   &$2.13\pm 0.03$   &$1.82\pm 0.08$   &$77_{-43} ^{ \dagger }$   &${4.3_{-0.2}^{+1.1}}$
%& $10^{4.3_{-0.2}^{+1.1}}$
&19.00$^{+0.06} _{-0.02}$
%&  $10^{19.0^{+0.06}_{-0.02}}$ 
\rule[-1.5mm]{0mm}{5mm}\\
Wing 1   &$4.29\pm 0.23$   &$12.55\pm 0.63$   &$12.00\pm 0.60$   &$33_{-3}^{+11}$   &${9.0_{-4.4}^{ \dagger }}$
%& $10^{9.0_{-4.4}^{ \dagger }}$
&20.00$^{+0.02} _{-0.04}$
%&  $10^{20.0^{+0.02}_{-0.04}}$
\rule[-1.5mm]{0mm}{5mm}\\\hline
\end{tabular}
\footnotetext[0]{
$^{\dagger}$The upper bounds could not be found within the parameter ranges of our calculations:
\\
\quad \, $10\,\mbox{[K]}\!\leq\!T_{\rm k}\!\leq\!100\,\mbox{[K]}$ and $10^{1}\,[\mbox{cm}^{-3}]\!\leq\!n({\rm H}_2)\!\leq\!10^{9}\,[\mbox{cm}^{-3}]$.
}
\end{minipage}
\end{table*} %\footnotemark[1] 
%%%%%%%%%%%%%%%%%%%%%%%%%%%%%%%%%%%%%%%%%%%%%%%%%%%%%%%%%%%%%%%%%%%%%%%%%%%%%%%%%%%%%%%%%%%%%%%%%%%%%%%%%%%%%%%%%

Using the physical conditions derived for the Bullet and assuming the optically thin limit, we estimated the total mass of the Bullet from CO {\it J}=3--2 intensity to be $7.5 \,M_{\sun}$. If the Bullet arises from the W44 molecular cloud at $\VLSR\!\simeq\!40 \ \mbox{\kms}$, its kinetic energy amounts to $10^{48.0}  \ \mbox{\erg}$. This is approximately 1.5 orders of magnitude greater than the kinetic energy shared to the small solid angle of the Bullet, $(1\mbox{--}3)\!\times\!10^{50}\ \mbox{\erg} \,\times (\Delta\Omega/4\pi )\!=\!10^{46.3\mbox{--}46.8} \ \mbox{\erg}$.   
Obviously, the energy injection by an SN explosion alone cannot account for the kinetic energy of the Bullet.  

%%%%%%%    Place Fig.3 here, please.      %%%%%%% 
\begin{figure}[htbp]
\epsscale{1}
\plotone{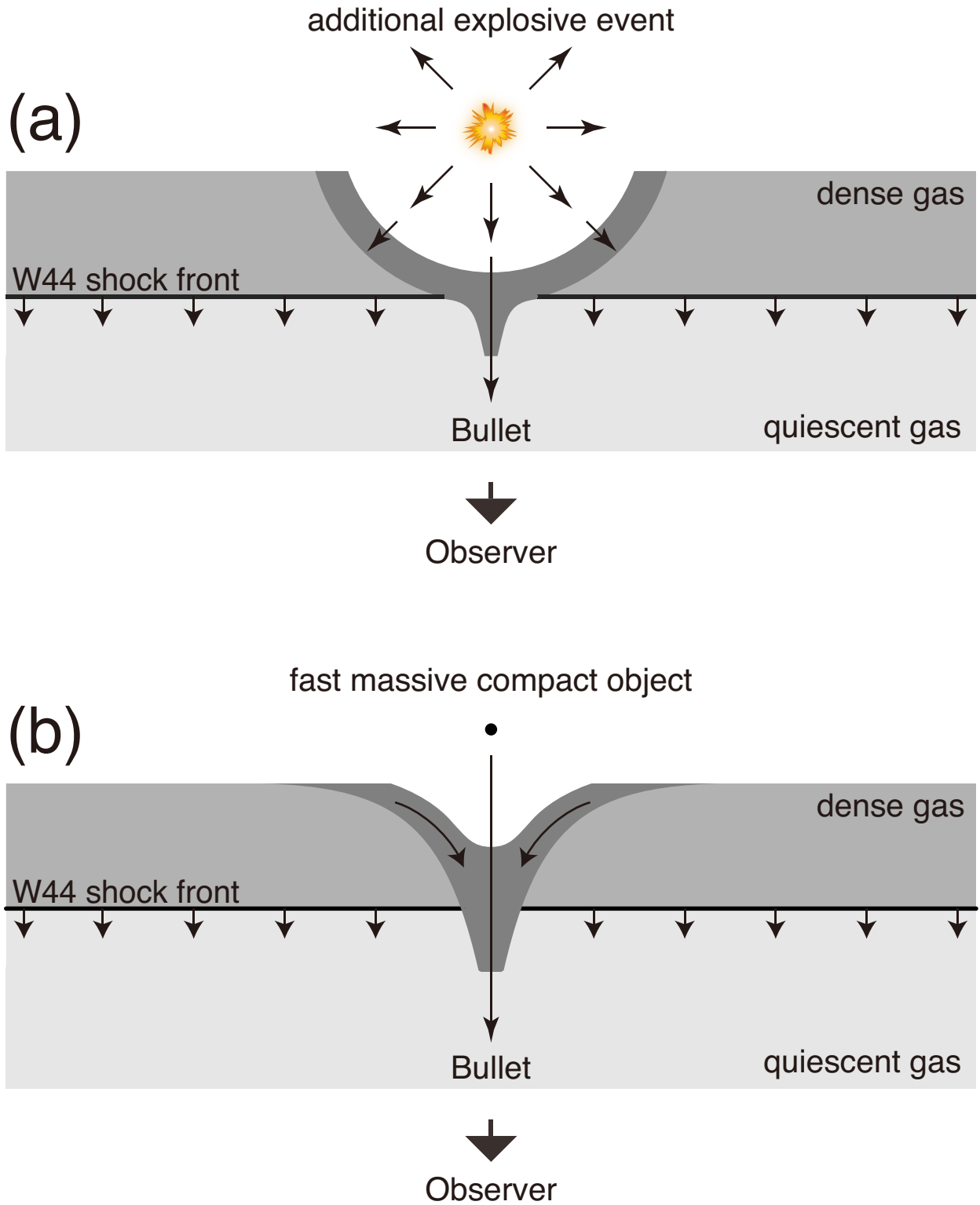}
\caption{Schematic views of the formation scenario of the Bullet; (a) the explosion model and (b) the shooting model. Line of sight is the direction from bottom to top.}
\label{fig3}
\end{figure}  
%%%%%%%%%%%%%%%%%%%%%%%%%%%%%%

\subsection{Origin of the Bullet}
The characteristics of the Bullet are summarized as follows: 
\begin{enumerate}
\item It has a compact appearance ($0.5\times 0.8$ pc$^2$) and an extraordinarily broad velocity width ($100 \ \mbox{\kms}$), giving a dynamical time of 5000--8000 yr.  
\item Unlike previously detected broad emissions, it appears in the negative velocity side only.
\item It bifurcates at $\VLSR\simeq -10 \ \mbox{\kms}$, showing a ``Y'' shape in the {\it l-V} plane.  
\item It has a mass of $7.5$ $M_{\sun}$ and a kinetic energy of $10^{48.0}  \ \mbox{\erg}$. The kinetic energy cannot be supplied by the W44 SN shell.  
\item A radio continuum blob and an H$_2$ ro-vibrational line nebulosity are associated with it.  
\end{enumerate}
These characteristics indicate that the Bullet has been accelerated by a certain local event, possibly driven by a single source.  If so, the kinetic energy supplied by the driving source must be greater than $10^{48.0}  \ \mbox{\erg}$.  In addition, the short dynamical time and large kinetic energy indicate that the kinetic power of the driving source must be greater than $10^{36.6} \ \mbox{\ergs}$ ($\simeq 1000$ $L_{\sun}$).  
Here, we present two possible formation scenarios of the Bullet.

\subsubsection{Explosion Model}
An additional explosive event near the expanding SN shell could account for the kinematics of the Bullet (Fig. 3a).  In this case, the driving source must be located behind the expanding shell since the Bullet is single-sided.  For the same reason, the kinetic energy and kinetic power necessary for the Bullet formation are doubled, $10^{48.3} \ \mbox{\erg}$ \ and $10^{36.9} \ \mbox{\ergs}$, respectively.  The upper part of the ``Y'' shape can be interpreted as the expanding shell due to the local explosion, while the lower part might be the broken tip of the shell in the low-density layer.  

A rotation-powered pulsar may be a candidate for the driving source.  Some energetic pulsars have spin-down power higher than $10^{37} \ \mbox{\ergs}$ (e.g., Arumugasamy, Pavlov, \&\ Kargaltsev 2014). However, no pulsar has been detected toward this position.  The W44 pulsar is located at least $\sim\!8$ pc away from the Bullet.  A bipolar outflow from a massive protostar is not a candidate for the driving source, because of the absence of a luminous stellar object, although some of them provide kinetic energies greater than $10^{48.0} \ \mbox{\erg}$ (Bally \&\ Zinnecker 2005).  

%{\bf
A completely separate supernova explosion can account for the kinetic energy of the Bullet.  
If we assume an isotropic disk (radius$=\!15$ kpc, thickness$=\!200$ pc) distribution of population I objects and that supernovae occur in spiral arms (volume filling factor$\sim\!0.3$), the supernova rate of the entire Galaxy ($\sim\!0.02$ yr$^{-1}$; Tammann et al. $1994$) provides a chance probability of another supernova in the W44 shell (radius$=\!15$ pc) within 20 thousand years as $\sim\!10^{-3.8}$.  Considering that massive stars are formed in clusters, this chance probablity may be further enhanced.  

Despite the separate supernova scenario can not be ruled out, here we consider another possibility in detail.  That is an isolated black hole (BH) activated by the passage of a high-density molecular gas layer.  
%}
Such an object has already been suggested as a driving source of the Tornado nebula (Sakai et al. 2014). This process is formalized as the Bondi-Hoyle-Littleton (BHL) accretion (e.g., Edgar 2004). Assuming the standard accretion disk model onto a BH and that all the gravitational energy is converted to the kinetic energy, the BHL accretion with a duration time $\Delta t$ provides 
\begin{eqnarray}
E_{\rm HL} & = &\frac{1}{12} \dot{M}_{\rm HL} c^2 \Delta t \\
\dot{M}_{\rm HL} & = &  \frac{4\pi G^2 M^2 \rho}{(v^2+\sigma^2)^{3/2}} \ ,
\end{eqnarray}
where $M$ is the BH mass, $\rho$ is the mass density of the ambient interstellar medium, $v$ is the relative velocity, and $\sigma$ is the velocity dispersion. The duration time of the accretion is calculated by $\Delta t = l/v$, where $l$ is the thickness of the high-density layer. If we take $l=0.1 \ \mbox{pc}$, which is about the thickness of thin filaments detected in W44 at 20 cm continuum emission (Jones et al 1993), we have $\Delta t = 7500$ yr. Employing \mbox{$n({\rm H}_2)\!=\!10^{4.5}$ cm$^{-3}$}, \mbox{$v\!=\!13.2 \ \mbox{\kms}$}, \mbox{$\sigma\!=\!6.8 \ \mbox{\kms}$}  (Sashida et al. 2013), and the mean molecular weight of 1.36, we have 
\begin{equation}
E_{\rm HL} = 10^{48.3} \left( \frac{M}{3.4 \,M_{\sun}} \right)^2 \ \mbox{erg} .  
\end{equation}
This means a BH mass of greater than 3.4 $M_{\sun}$ could account for the kinetic energy of the Bullet. The putative BH is currently inactive since the accretion has already ceased.

%%%%%%%    Place Fig.4 here, please.      %%%%%%% 
\begin{figure}[htbp]
\epsscale{1}
\plotone{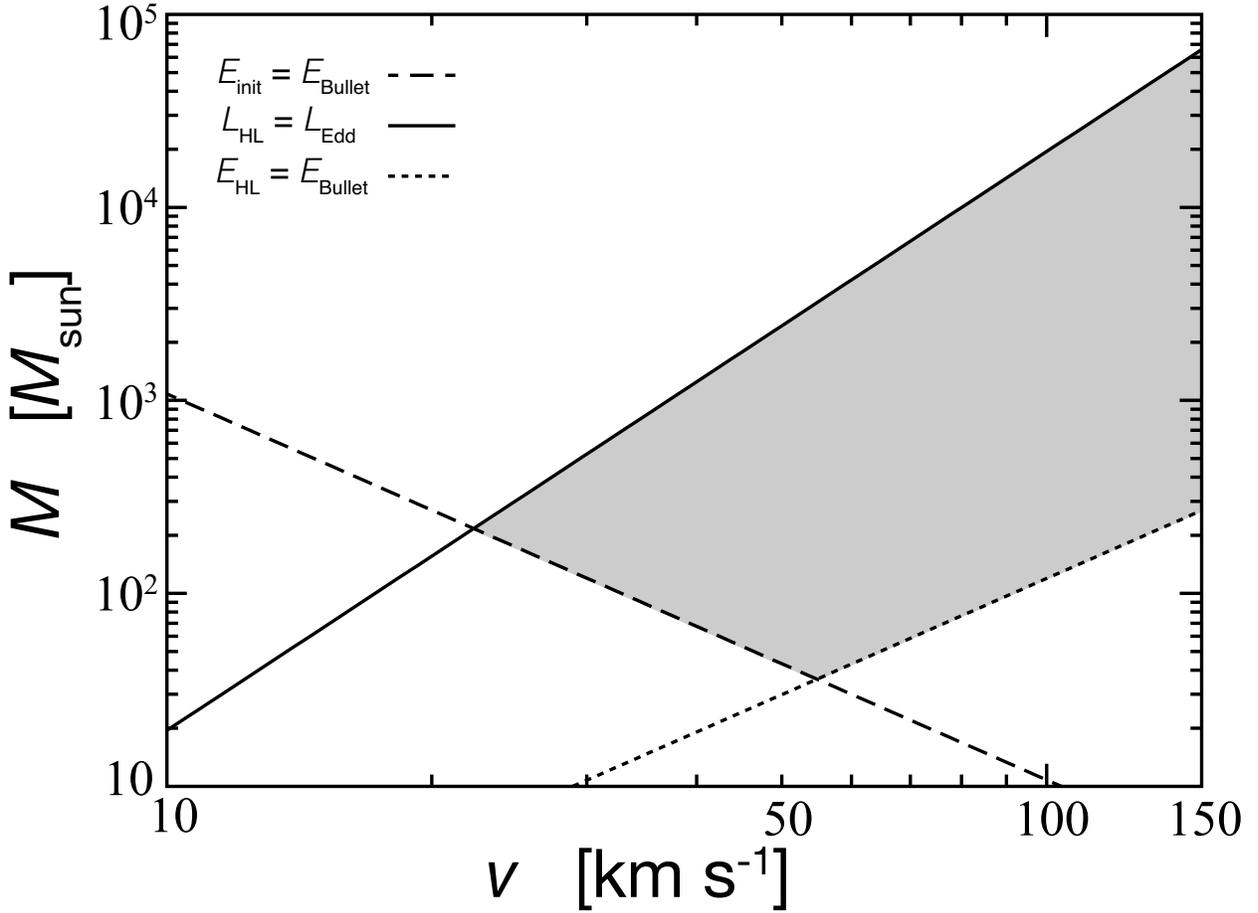}
\caption{Lines of $E_{\rm init}\!=\!10^{48.0}  \ \mbox{\erg}$, $E_{\rm HL}\!=\!10^{48.0}  \ \mbox{\erg}$, and $L_{\rm HL}\!=\!L_{\rm Edd}$ in the $M$-$v$ plane. Gray shaded area indicates the permitted range of parameters.  
\label{fig4}}
\end{figure}  
%%%%%%%%%%%%%%%%%%%%%%%%%%%%%%

\subsubsection{Shooting Model}
Conversely, a plunge of a high-velocity massive object into the high-density layer toward us could also explain the kinematics of the Bullet (Fig. 3b). This process can be described by the BHL accretion formula as well. The upper part of the ``Y'' shape can be interpreted as the shock wave propagating in the high-density layer, while the lower part might correspond to the ``wake'' accreting to the plunging object (e.g., Edgar et al. 2004). Again, the absence of a luminous stellar object favors a drifting, isolated BH as a candidate for the plunging object.   

The plunging velocity must be higher than $13.2$ \kms\ but it is not necessarily as high as $100 \ \mbox{\kms}$\ since gas in the ``wake'' is accelerating toward the object. Of course, the initial kinetic energy of the plunging object, \mbox{$E_{\rm init}\!=\!(1/2) M v^2$}, must be greater than \mbox{$E_{\rm Bullet}\!=\!10^{48.0}  \ \mbox{\erg}$.} Setting $v$ to the plunging velocity, we can calculate the total energy available from the BHL accretion by using equation (1). This BHL energy $E_{\rm HL}$ must also be greater than $E_{\rm Bullet}$. In addition, the BHL luminosity, $L_{\rm HL}\!=\!(1/12) \dot{M}_{\rm HL} c^2$, probably does not exceed the Eddington limit ($L_{\rm Edd}$). Figure 4 shows the range of $(M, v)$ permitted by the above conditions, indicating that the plunging object must have a mass greater than $30\,M_{\sun}$.  

This shooting model, as well as the explosion model, can reproduce the broad velocity width and the enormous kinetic energy of the Bullet. A naive shooting model predicts that the size of the accreting zone should be of the same order as the Hoyle--Litteleton radius, $r_{\rm HL}\!=\!2GM/(v^2+\sigma^2) \sim 10^{-5} $ pc.
Since the observed size of the negative high-velocity tail of the \mbox{Bullet ($\sim 0.3$ pc)} is comparable to the telescope beamsize (22\arcsec), the real spatial size might be much smaller. Moreover, the magneto-hydrodynamical effect can enlarge the accreting zone, which is roughly estimated by $(v_{\rm A}/v)\,l$, where $v_{\rm A}$ is the Alfv\'en velocity and $l$ is the thickness of the high-density layer (Nomura et al. 2016 in prep.). The magnetic field strength should be enhanced in the shocked gas layer, and thereby increases $v_{\rm A}$ and the size of the accreting zone accordingly.

\subsection{Implications}
Either scenario includes an isolated BH, which is currently inactive. But is it reasonable?  Actually, there must be a number of invisible BHs in the galaxy.  For example, Agol \&\ Kamionkowski (2002) estimated the total number of stellar mass BHs in the galaxy to be $\sim\!10^9$, which is far greater than the latest total number of the Galactic BH candidates ($\sim\!60$).  Assuming a spherically isotropic distribution with a radius of 30 kpc, this total BH number corresponds to a number density of $10^{-5.1}$ pc$^{-3}$.  Thus, we obtain the expected value of the BH number within the W44 shell (15 pc radius) as 0.13, which is not significantly low. Note that this value may be a lower limit since an overdensity of BHs due to the disk population and a %`fossil' 
OB association ($\rm \ddot{O}$gelman et al. 1976)
%(e.g., Knapp et al. 1974) 
is expected.

Considering the permitted mass ranges of an isolated BH, the explosion model seems to be more plausible.  Such an isolated, currently inactive BH with a mass \mbox{of $>$ 10 $M_{\sun}$} was suggested as the driving source of the Tornado nebula (Sakai et al. 2014).  We also detected another extremely broad-velocity-width feature, CO--0.40--0.22, in the central molecular zone of our Galaxy (Oka et al. 2016).  This was explained by a gravitational kick to the molecular cloud caused by an intermediate-mass BH.  Therefore, an elaborate search for compact high-velocity features using molecular lines might be an effective method of seeking inactive BHs in the Galaxy and nearby galaxies.

\acknowledgments
This paper is based on observations conducted using the Nobeyama Radio Observatory (NRO) 45-m telescope and ASTE. The NRO is a branch of the National Astronomical Observatory of Japan (NAOJ), National Institutes of Natural Sciences. We are grateful to the NRO staff and all the members of the ASTE team for operation of the telescope. T.O. acknowledges support from JSPS Grant-in-Aid for Scientific Research (B) No. 15H03643.

\end{document}